\documentstyle[11pt,epsfig,rotate]{article}
\begin{document}
\def\Tr{\,{\rm Tr}\,}
\def\beq{\begin{equation}}
\def\eeq{\end{equation}}
\def\beqa{\begin{eqnarray}}
\def\eeqa{\end{eqnarray}}
\begin{titlepage}
\vspace*{-1cm}
\noindent
\phantom{bla}
\hfill{$\scriptstyle{\rm UMHEP-453}$}
\\
\vskip 2.0cm
\begin{center}
{\Large {\bf The leading chiral electromagnetic correction to the
nonleptonic $\Delta I = 3/2$ amplitude in kaon
decays}}
\end{center}
\vskip 1.5cm
\begin{center}
{\large Vincenzo Cirigliano$^a$, John F. Donoghue$^b$ 
and Eugene Golowich$^b$} \\
\vskip .15cm
$^a$ Dipartimento di Fisica dell'Universit\`a and I.N.F.N. \\
Via Buonarroti,2 56100 Pisa (Italy) \\
vincenzo@het.phast.umass.edu \\

\vskip .15cm

$^b$ Department of Physics and Astronomy \\
University of Massachusetts \\
Amherst MA 01003 USA\\
donoghue@het.phast.umass.edu \\
gene@het.phast.umass.edu \\

\vskip .3cm
\end{center}
\vskip 1.5cm
\begin{abstract}
\noindent
In kaon decay, electromagnetic radiative corrections can
generate shifts in the apparent $\Delta I = 3/2$ amplitude
of order $\alpha A_0/A_2 \sim 22\alpha$. In order to know the true
$\Delta I = 3/2$ amplitude for comparison with lattice calculations
and phenomenology, one needs to subtract off this electromagnetic effect.
We provide a careful estimate of the leading electromagnetic shift 
in the chiral expansion of the 
amplitude, which shows that it is smaller than naive expectations, 
with a fractional shift of 
$\delta A_2^{\rm (em)} /A_2 = - 0.016 \pm 0.01$.
\end{abstract}
\vfill
\end{titlepage}

\section{\bf Introduction}
The $\Delta I = 1/2$ enhancement of nonleptonic kaon decays is a
well-known feature which has still not been completely explained. Most 
existing lattice calculations find a $\Delta I = 3/2$ amplitude
which is a factor of two larger than the experimental
value.\footnote{A recent lattice calculation is more promising in this
regard.~\cite{jlqcd}} This $\Delta I = 3/2$ amplitude also enters 
present phenomenology through the chiral determination of the $B_K$ 
parameter which is part of the Standard Model prediction of CP 
violation. The chiral calculation~\cite{dgh0} (which relates $B_K$ 
to the {\it experimental} $\Delta I = 3/2$ amplitude)  
disagrees with the quenched lattice calculation for $B_K$ also by
about a factor of two. At present we do not know if the disagreement 
in $B_K$ is due to the failure of the lattice approach to reproduce the 
experimental amplitude or if there are large chiral corrections 
responsible for the difference.

In both these applications, we need to know the {\it true} $\Delta I =
3/2$ effect. Since the $\Delta I= 1/2$ amplitude is so much larger, 
it is possible that electromagnetic corrections to it may simulate an
effect which is similar to the small $\Delta I=3/2$ amplitude. These
electromagnetic corrections enter at order $\alpha A_0 =
A_0/137$, which can be comparable to a sizeable portion 
of the $\Delta I=3/2$ amplitude $A_2 \simeq A_0/22$. The possibility 
then emerges that the relevant experimental amplitude 
(without electromagnetism) could differ significantly from 
that presently being used in phenomenology. 

Although the literature on this subject extends over many 
years, {\it e.g.} see Refs.~\cite{aly,dbm,ns,bk}, the issue 
has not yet received a definitive treatment. It is our aim 
in this paper to provide an analysis using the most 
up-to-date tools which will yield a reliable estimate 
of this effect.  In a longer paper, we will present a more 
comprehensive analysis of
electromagnetic radiative corrections in kaon decays. The full system,
particularly the decay $K_S \to \pi^+ \pi^-$, brings in several
additional complications, such as the Coulomb effect on the final
state, the violations of Watson's theorem from the mixing of final
states and the induced $\Delta I = 5/2 $ effect. However, the decay
$K^+ \to \pi^+\pi^0$ is particularly simple and can by itself be used
to answer the question that we have raised above. Among our results 
in this paper are: 
\begin{enumerate}
\item The long distance portion of the leading chiral result 
satisfies the relation ${\cal M}_{\rm LD} = - 2 g_8 (\delta 
m_\pi^2)_{\rm LD} /F_\pi^2$ where $(\delta m_\pi^2)_{\rm LD}$ 
is the long distance part of the pion electromagnetic 
mass difference and the other constants are defined below.
\item This leading long distance effect is canceled exactly 
by the effect of pion mass differences in the usual weak amplitude.
\item There are, however, residual effects coming from intermediate 
energies and the electroweak penguin operator. Although we allow
generous uncertainties associated with intermediate energies, it is
clear that the net residual effect is quite small.
\end{enumerate}

\section{Chiral Analysis}
Chiral symmetry provides the framework for structuring our 
calculation. We first replace the calculation of 
the decay amplitude $A_{+0}(p_K,p_{\pi^+},p_{\pi^0})$ with 
the simpler $K^+$-to-$\pi^+$ matrix element by taking 
the soft pion limit, 
\beq
A_{+0}(p,p,0) = - {i \over 2 F_\pi} {\cal M}(p) \ \ , 
\label{2}
\eeq
where 
\beq
_{\rm out}\langle \pi^+ (p') | K^+ (p) \rangle_{\rm in}
= i (2 \pi)^4 \delta^{(4)} (p' - p) \ \ {\cal M}(p) \ \ .
\label{1}
\eeq
For most of our analysis, we shall work with the leading chiral 
component ${\cal M}(0)$ 
\beq
{\cal M}(p) \ = \ {\cal M}(0)\ + \ {\cal O}(p^2) + \dots \ \ ,
\label{leading}
\eeq
and add in (small) ${\cal O}(p^2)$ contributions at the end.   
\begin{figure}
\vskip .1cm
\hskip 1.5cm
\epsfig{figure=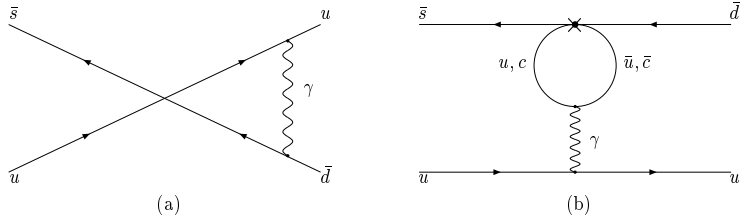,height=1.1in}
\caption{Combined weak-electromagnetic transitions 
of quarks.\hfill 
\label{fig:fig1}}
\end{figure}

\subsection{The Chiral and Electromagnetic-penguin Components}
At the quark level, electromagnetic corrections 
to the weak transition ${\bar s}+u \to {\bar d}+u$ fall 
into two distinct classes ({\it cf} Fig.~1), which we call 
the chiral (CH) and electromagnetic-penguin (EMP) components, 
\beq
{\cal M}(0) = {\cal M}_{\rm CH} + {\cal M}_{\rm EMP} \ \ .
\label{define}
\eeq

The amplitude ${\cal M}_{\rm CH}$ is associated with  
the long-range and intermediate-range contributions of 
the process in Fig.~1(a) (along with all other 
diagrams in which a photon is exchanged between 
the quark legs).  We note that the short distance part of 
such transitions leads merely 
to an overall shift in the strength of the nonleptonic 
interaction. Using the procedure described in Ref.~\cite{dgh1}, 
this is equivalent to an effective Fermi constant defined as 
\beq
G_{NL} ( {\bar \mu} ) = G_\mu \left[ 1+{2 \alpha\over 3\pi} 
\ln\left({M_W \over {\bar \mu}}\right)\right] \ \ ,
\label{sd1}
\eeq
where ${\bar \mu}$ is an energy scale lying in the region 
where perturbation theory is valid 
and $G_\mu$ is the Fermi constant measured in muon decay. 
This shift does not lead to mixing of isospin 
amplitudes and is irrelevant for the purposes of this paper.
Calculation of the long and intermediate range contributions 
to ${\cal M}_{\rm CH}$ is carried out in Sect.~3.1 
and Sect.~3.2.  

The amplitude ${\cal M}_{\rm EMP}$ corresponding to the 
electromagnetic penguin operator of Fig.~1(b) will have 
both long-distance and short-distance components.  These 
are described in Sect.~3.3.  Determination of the full 
amplitude ${\cal M}$ is carried out in Sect.~3.4, with 
special attention paid to the relative phase between 
${\cal M}_{\rm CH}$ and ${\cal M}_{\rm EMP}$ and to the matching of
long and short distances. 

\subsection{Chiral Lagrangians}
We now introduce some useful chiral lagrangians.  
The weak interactions involve left-handed currents only and the 
nonleptonic hamiltonian has an octet and 27-plet component. The 
lowest-order weak lagrangian for the dominant octet 
portion involves two derivatives,
\beq
{\cal L}_8 \ = \ g_8 \Tr \left( \lambda_6 D_\mu U D^\mu U^\dagger
\right) \ \ ,
\label{12}
\eeq
with $|g_8| \simeq 7.8 \cdot 10^{-8}~F_\pi^2$.  

Electromagnetic corrections involve both left-handed and 
right-handed effects, and can lead to lagrangians which do not 
involve derivatives.  For example, one of the effects
of electromagnetism is to shift the charged pion masses, an effect 
described at lowest order by the lagrangian
\beq
{\cal L}_{\rm ems} = g_{\rm ems}
\Tr \left( Q U Q U^\dagger \right) \ \ .
\label{98}
\eeq
The parameter $g_{\rm ems}$ is fixed from the pion 
electromagnetic mass splitting, 
\beq
\delta m_\pi^2 \ = \ {2 \over F_\pi^2} g_{\rm ems}  \ \ .
\label{98a}
\eeq

There is also the lagrangian which describes the leading 
electromagnetic correction to the weak interactions to leading 
chiral order,\footnote{We also make use of lagrangians 
which couple pions and kaons to resonances, as in 
Figs.~2(b),3(b).} 
\beq
{\cal L}_{\rm emw} = g_{\rm emw}
\Tr \left( \lambda_6 U Q U^\dagger \right) \ \ , 
\label{99}
\eeq
where $g_{\rm emw}$ is an {\it a priori} unknown coupling 
constant.  Knowledge of $g_{\rm emw}$ is equivalent 
to that of the matrix element ${\cal M}(0)$ 
as the two are related in the chiral limit by 
\beq
{\cal M}(0) \ = \ {2 \over F_\pi^2}~ g_{\rm emw} \ \ .
\label{99a}
\eeq

In Sect.~3, we present a detailed calculation 
of ${\cal M}(0)$ (and thus of $g_{\rm emw}$) 
and as a consequence reveal an approximate numerical 
relationship between $g_{\rm emw}$ and $g_{\rm ems}$.  
Then in Sect.~4, we turn to the full 
$A_{+0}$ amplitude, including also the effects that arise 
at the next chiral order (${\cal O}(p^2)$) from the photon loop
calculation. 

\section{\bf Calculation of the leading chiral amplitude}
To fully calculate the relevant amplitude, we need to consider
contributions from all scales. We will recognize three regions of the
virtual photon momentum: 
\begin{enumerate}
\item very low energies $Q^2 < \Lambda^2$ with $\Lambda \sim m_\rho$, 
\item high energies with $Q^2 > \mu^2$ \ \  
($\mu \sim 1.5 \to 2.5~{\rm GeV}$), 
\item intermediate energies between these two regions. 
\end{enumerate}
In the lowest energy regime, we may use chiral techniques to obtain 
the leading effect. At high energies, the short distance analysis 
of QCD will be employed. The most important ingredient of the
treatment of intermediate energies 
is the requirement of matching these
two regions. This will be modelled on 
physics which is reliably known in the case of electromagnetic mass shifts.

\begin{figure}
\vskip .1cm
\hskip 8.5cm
\epsfig{figure=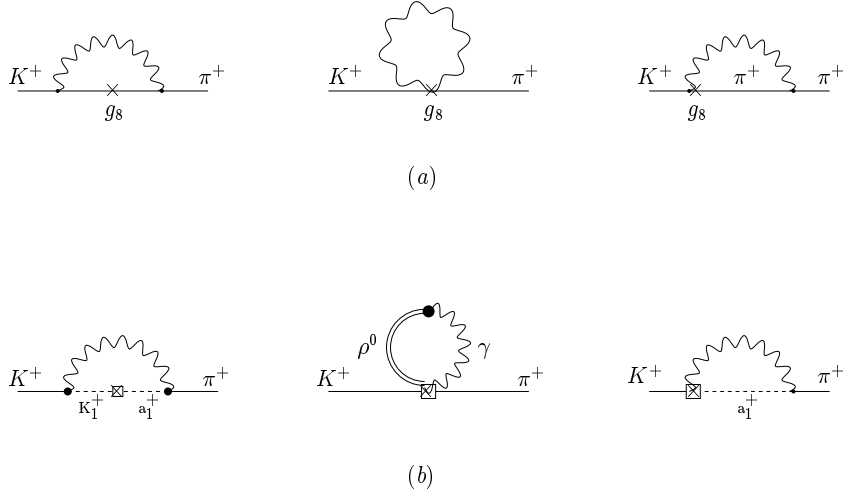,height=2.55in}
\caption{Electromagnetic corrections to the 
$K^+$-to-$\pi^+$ transition. \hfill 
\label{fig:fig2}}
\end{figure}

\subsection{\bf Long Distance Component of ${\cal M}_{\rm CH}$}
The very long distance component can be calculated from the 
combined chiral lagrangian of the strong weak and electromagnetic 
interactions in the chiral limit. In the diagrams of Fig.~2(a), 
one finds after Wick rotation a matrix element
\beq
{\cal M}_{\rm LD} \ = \ - { 3 \alpha g_8\over 2\pi F_\pi^2}
\int_0^{\Lambda^2} dQ^2 \ \ , 
\label{ld1}
\eeq
where $\Lambda$ represents the upper end of the low-energy 
region.  We note that a similar calculation of the electromagnetic 
mass shift of the charged pion, Fig.~3(a), yields
\beq
\delta m_\pi^2 \Big|_{\rm LD} \ =\  {3 \alpha \over 4\pi}
\int_0^{\Lambda^2} dQ^2 \ \ ,
\label{8}
\eeq
The similarity of the two can be motivated by the fact that in 
the former calculation the weak vertex in the loop introduces a
factor of $l^2$ ($l$ is the loop momentum) which compensates one of
the two propagators, yielding an effect similar to that of Fig.~3(a).
At this stage, it is a curiosity to note that the choice of $\Lambda^2
= m_\rho^2$ provides an accurate description of the pion mass
difference. However, we will see below that this is not an 
accident --- 
that reliably known physics cuts off the integral above the rho mass.
For our purposes at this stage, this similarity is the first
indication of the relation 
\beq
{\cal M}_{\rm LD} 
= - 2 {g_8\over F_\pi^2} (\delta m_\pi^2)_{\rm LD}  
\label{ld3}
\eeq
or $(g_{\rm emw})_{\rm LD} = - g_8 {\delta m_\pi^2}_{\rm LD}$.
This could be derived somewhat more formally by using a rotation to the
basis where the kinetic energy matrix is diagonalized.~\cite{rotated} 
The application of long distance electromagnetic mass shifts to the
rotated basis is equivalent to the above relation in the non-rotated
basis. 

\begin{figure}
\vskip .1cm
\hskip 5.cm
\epsfig{figure=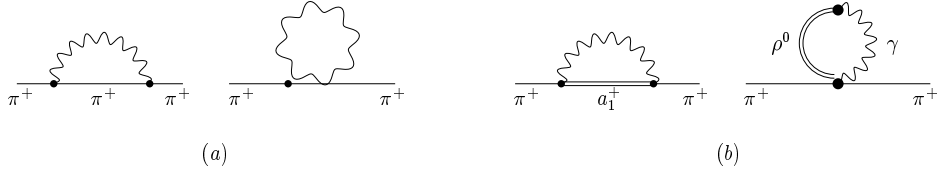,height=0.85in}
\caption{Pion electromagnetic mass shift.\hfill 
\label{fig:fig3}}
\end{figure}

\subsection{\bf Intermediate Energy Component of ${\cal M}_{\rm CH}$} 
A prototype for dealing with the intermediate energies is 
the pion electromagnetic mass difference. The accumulated 
wisdom of many studies 
has given us an accurate guide to the physics of this process. A 
rigorous approach would involve the sum rule of Das 
{\it et al}~\cite{dgmly}, in which
$\delta m_\pi^2$ is expressed in terms of the difference of the 
experimental vector and
axialvector spectral functions $(\rho_V - \rho_A)(s)$. This has
been analysed successfully using experimental data and QCD
constraints.~\cite{dg1} A simplified expression that captures the essential
physics is obtained upon saturating $\rho_V$ and $\rho_A$ 
respectively with the vector $\rho$ resonance and the axialvector 
resonance $a_1$.  This yields 
\beqa
\delta m_\pi^2  \ & = &\   { 3 \alpha \over 4\pi } 
\int_0^{\infty}
dQ^2 \left[ 1 - {F_V^2 \over F_\pi^2}{Q^2\over Q^2 + m_\rho^2} +
{F_A^2 \over F_\pi^2} {Q^2\over Q^2 + m_{a_1}^2}
\right] \nonumber \\
 & = & {3 \alpha \over 4\pi}
\int_0^{\infty} dQ^2\ {m_{a_1}^2 \over Q^2 + m_{a_1}^2}  \cdot
{m_\rho^2 \over Q^2 + m_\rho^2} \ \ .
\label{11}
\eeqa
The second form is found when the resonance couplings and masses
satisfy the Weinberg sum rules, which are required in order to obtain
the right high energy behavior.
The long-distance amplitude given in Eq.~(\ref{8}) has been softened at values
of $Q^2$ above the meson masses so that $m_\rho$ and $m_{a_1}$ act as the
effective cutoff for the integral. This result is equally well
reproduced by introducing resonance couplings to the effective
lagrangian and imposing the Weinberg sum rules on the masses and
couplings. This involves the diagrams of Fig.~3(b).

What is the analogous statement for ${\cal M}_{\rm CH} \equiv 
{\cal M}_{\rm LD} + {\cal M}_{\rm INT} + {\cal M}_{\rm SD}$?  
First, we recall from the discussion surrounding 
Eq.~(\ref{sd1}) that, rather than contributing to the isospin 
mixing effect, the dominant effect of the short distance 
(SD) component is to renormalize the Fermi constant.   
The full chiral amplitude thus experiences important 
contributions only from long and intermediate distance effects and
must vanish in the short distance region.  
Combining results from the previous sections, our form for the long
and intermediate distance regions is 
\beq
{\cal M}_{\rm CH} 
= - { 3 \alpha g_8\over 2\pi F_\pi^2} 
\int_0^{\mu^2}
dQ^2 \left[ 1 - {B_V Q^2\over Q^2 + m_V^2} -
{B_A Q^2\over Q^2 + m_A^2}
+ {C m_A^2Q^2 \over (Q^2 + m_A^2)^2}
\right] \ \ ,
\label{13}
\eeq
where the $Q^2$-integral for ${\cal M}_{\rm CH}$ is seen to be 
effectively cut off at some scale $\mu^2$.  The quantities 
$B_V$, $B_A$ and $C$ in the above contain couplings from the 
weak interaction resonance lagrangians.~\cite{ekw}  It is of 
course possible to model these couplings, and we have done so.  
However, it is more to the point to implement the constraint 
(discussed earlier) that ${\cal M}_{\rm CH}$ receive no mixing 
contribution from the high-$Q^2$ region. Thus these constants must be
constrained such that the ampliude vanish at the matching to the short
distance region. For this to occur, 
we require that the large $Q^2$ limit 
of the integrand ($Q^2 \gg m_{V,A}^2$) approach zero rather than a
constant, i.e.
\beq
B_V \ + \ B_A \ = \ 1 \ \ .
\label{i4}
\eeq
We further constrain this amplidude by choosing the matching scale
$Q=\mu$ at which the amplitude vanish.
If $m_V$ and $m_A$ were equal this constraint would determine 
unknown $C$ in the integrand of Eq.~(\ref{13}).  
In our numerical study, we choose $\mu$ to lie 
between $1.5$~GeV and $2.5$~GeV and treat the resulting variation 
as one of the uncertainties of the calculation. 
Of course, the difference between $m_V$ and $m_A$ leads to a 
slight further uncertainty. To model this effect, we 
have explored models for the resonance couplings~\cite{weakres} 
|which weight the axialvector and vector resonances differently.  
It is found that this uncertainty is smaller than that associated 
with variation of the matching scale $\mu$. 

\subsection{\bf Determination of ${\cal M}_{\rm EMP}$}
As we shall show in the following, 
the electromagnetic penguin (EMP) operator of Fig.~1(b) 
gives rise to contributions over all distance scales.  
It is, however, the sole source of meaningful short distance 
effects in our calculation of electromagnetic corrections 
to $A_2$.  If the numerically tiny $t$-quark contribution is omitted, 
the EMP hamiltonian takes the form, 
\beqa
& & -i {\cal H}_{\rm EMP} = {\bar G} \int {d^4 q \over (2 \pi)^4} 
{I^{\mu\nu}(q,{\bar \mu}) \over q^2 + i \epsilon} 
\nonumber \\
& & \phantom{xxx} \times \int d^4 y \ 
e^{-i q \cdot y}~ T\left[ {\bar s}(0) \gamma_\mu (1 + \gamma_5) d(0) 
{\bar q}(y) Q \gamma_\nu q(y)\right] \ \ ,
\label{emp1}
\eeqa
where $q = u,d,s$ is a light-quark field, 
${\bar G} \equiv 2e^2 G_F V_{us}^* V_{ud}/(3\sqrt{2})$, 
and $I^{\mu\nu}(q,{\bar \mu})$ represents the effect of the 
quark-antiquark loop in the EMP operator.  In evaluating 
$I^{\mu\nu}(q,{\bar \mu})$, it is understood that at the lower end, 
the loop momentum is cut off at scale ${\bar \mu}$ ({\it cf} see 
Eq.~(\ref{emp7})).  The dependence on ${\bar \mu}$ is 
logarithmic and thus quite weak.  

\begin{figure}
\vskip .1cm
\hskip 3cm
\epsfig{figure=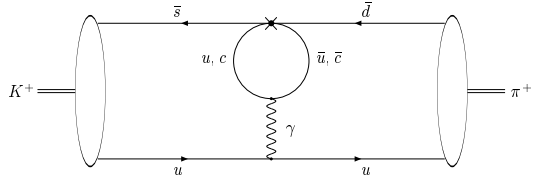,height=0.9in}
\caption{Electromagnetic penguin and the 
$K^+$-to-$\pi^+$ transition. \hfill 
\label{fig:fig4}}
\end{figure}

It can be shown~\cite{dg2} that 
in the chiral limit the $K^+$-to-$\pi^+$ matrix element of 
${\cal H}_{\rm EMP}$ ({\it cf} Fig.~4) is expressible as 
\beqa
& & \lim_{p=0}~ \langle \pi^+ (p) | T\left[ {\bar s}(0) \gamma_\mu 
(1 + \gamma_5) d(0) {\bar q}(y) Q \gamma_\nu q(y)\right]|K^+ (p) \rangle 
\nonumber \\
& & \phantom{xxxx} = -i {2 \over F_\pi^2} 
\int {d^4 k \over (2\pi)^4} \ e^{i k \cdot y} 
\left( k_\mu k_\nu - k^2 g_{\mu\nu} \right) 
\left[ \Pi_{\rm V3} - \Pi_{\rm A3} \right] (k^2)  \ \ , 
\label{emp3}
\eeqa
where $\Pi_{\rm V3}$, $\Pi_{\rm A3}$ are the isospin vector 
and axialvector correlators.  In the chiral limit, the 
$K^+$-to-$\pi^+$ matrix element of the EMP operator thus becomes 
\beqa
& & \lim_{p=0}~ \langle \pi^+ (p) | {\cal H}_{\rm EMP} 
|K^+ (p) \rangle 
\nonumber \\
& & \phantom{xxx} = {2 {\bar G} \over F_\pi^2} 
\int {d^4 q \over (2 \pi)^4} 
{I^{\mu\nu}(q,{\bar \mu}) \over q^2 + i \epsilon} 
\left( q_\mu q_\nu - q^2 g_{\mu\nu} \right) 
\left[ \Pi_{\rm V3} - \Pi_{\rm A3} \right] (q^2) \ \ .
\label{emp5}
\eeqa
This expression describes the EMP effect over all 
scales of the virtual photon (euclidean) momentum, 
\beq
{\cal M}_{\rm EMP} =
- {3 {\bar G} M_V^2 M_A^2 \over (2 \pi)^4} 
\int_0^\infty dQ^2 \ {Q^2 \over (Q^2 + M_V^2)(Q^2 + M_A^2)}
I (Q^2,{\bar \mu}^2) \ \ ,
\label{emp6}
\eeq
where we have expressed the $Q^2$ dependence of the 
correlators in terms of vector and axialvector pole terms, 
an approximation we know to be valid to within a few per cent.  
An explicit form for the EMP integral $I (Q^2,{\bar \mu}^2)$  is 
\beqa
I (Q^2,{\bar \mu}^2) &=& \int_0^1 dx \ x (1 - x) ~
\bigg[ \ln {{\bar \mu}^2 + m_c^2 + Q^2 x (1 - x)  \over 
{\bar \mu}^2 + m_u^2 + Q^2 x (1 - x) } 
\nonumber \\
& & + { m_c^2 + Q^2 x (1 - x)  \over  
{\bar \mu}^2 + m_c^2 + Q^2 x (1 - x)  }
- { m_u^2 + Q^2 x (1 - x)  \over  {\bar \mu}^2 + m_u^2 + Q^2 x (1 - x)  }
\bigg] \ \ .
\label{emp7}
\eeqa
where $m_c$ and $m_u$ are the $c$-quark and $u$-quark masses. 
The latter vanishes in the chiral limit.

\subsection{Matching}
The final step in determining ${\cal M}(0)$ is to add 
together the components ${\cal M}_{\rm CH}$ and ${\cal M}_{\rm EMP}$. 
This is displayed schematically in 
Fig.~5 which depicts the integrands in the $Q^2$ integrals and 
indicates that the short-distance contribution is numerically much 
smaller than the long-distance contribution.  

\begin{figure}
\vskip .1cm
\hskip 2.cm
\rotate[l]{\epsfig{figure=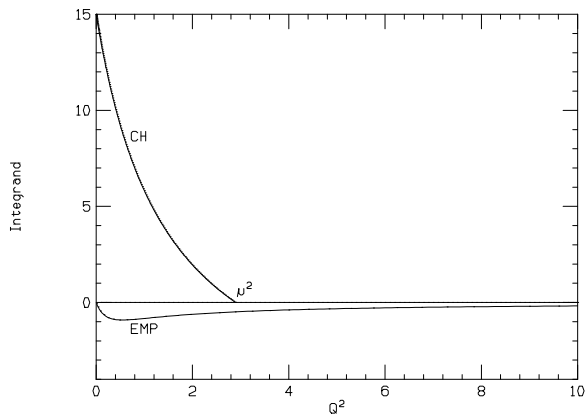,height=3.in}}
\caption{Schematic of matching. \hfill 
\label{fig:fig5}}
\end{figure}

It would appear that the calculation must 
contain an ambiguity arising from ignorance of the 
relative phase between ${\cal M}_{\rm CH}$ and ${\cal M}_{\rm EMP}$ 
or equivalentally of the sign of $g_8$.  However, we can infer 
that $g_8 < 0$ from the following argument.  From the 
$\Delta I = 1/2$ chiral lagrangian of Eq.~(\ref{12}), we have for 
the $K^0 \to (2\pi)_{I=0}$ amplitude, 
\beq
A_0 = i { \sqrt{2} \over F_\pi^3} \left( 
m_K^2 - m_\pi^2 \right) ~g_8 \ \ .
\label{m1}
\eeq
Alternatively, we can obtain $A_0$ using 
the effective lagrangian of the Standard Model 
\beq
{\cal L}_{\Delta S = 1} = - {G_F \over \sqrt{2}} 
V_{ud} V_{us}^* \sum_i \ c_i (\mu) {\cal O}_i \ \ ,
\label{m2}
\eeq
together with the vacuum saturation approximation (VSA), 
\beq
\langle (2\pi)_{I=0} | {\cal O}_i | K^0 \rangle =
B_i^{(0)} \langle (2\pi)_{I=0} | {\cal O}_i | K^0 
\rangle_{\rm VSA}  \ \ ,
\label{m3}
\eeq
to write 
\beq
A_0 = -i G_F F_\pi V_{ud} V_{us}^* \left( 
m_K^2 - m_\pi^2 \right)~ g_8^{\rm eff} \ \ ,
\label{m4}
\eeq
with 
\beqa
& & g_8^{\rm eff} = \sqrt{3 \over 2} \bigg[ 
- {1 \over 6} c_1 B_1^{(0)} + {5 \over 9} c_2 B_2^{(0)} 
+ {1 \over 3} c_3 B_3 + c_4 B_4 
\nonumber \\ 
& & \phantom{xxxxx} - 16 \left( { \langle {\bar q} q \rangle 
\over F_\pi^3} \right)^2  L_5
\left( {1 \over 3} c_5 B_5 + c_6 B_6 \right)
\bigg] \ \ .
\label{m5}
\eeqa
In the above, $L_5$ is a coefficient in the Gasser-Leutwyler 
${\cal O}(p^4)$ chiral lagrangian, the superscripts on $B_1^{(0)}$, 
$B_2^{(0)}$ signify $I=0$ for the final state $2\pi$ pair, 
and we refer the reader to Ref.~\cite{bfe} for further 
details.   Since (with the exception of $B_3$) the $\{ B_i \}$ are 
all positive, the $\{ |c_i| \}$ have been determined at NLO 
and specifically $c_1,c_6 < 0$, we conclude with reasonable 
certainty that $g_8 < 0$.  

The analysis described throughout this section then leads to 
the following value, expressed in units of $g_8 \delta m_\pi^2$, 
for the coupling $g_{\rm ewp}$ of Eq.~(\ref{99}),
\beq
{g_{\rm emw} \over g_8 \delta m_\pi^2} 
\ = \ -0.62 \pm 0.19 \ \ .
\label{100}
\eeq
The uncertainty arises almost entirely from variation of the 
parameter $\mu$ (we have set ${\bar \mu} = 1.5$~GeV in this 
determination).

\section{\bf The $K^+ \to \pi^+\pi^0$ transition}
We now turn to the physical $K^+ \to \pi^+\pi^0$ transition. 
This receives contributions from the true $\Delta I = 3/2$ 
interaction ($A_2^{\rm (true)}$), electromagnetic corrections 
($\delta A_2^{\rm (em)}$) 
and isospin-breaking effects ($\delta A_2^{\rm (iso-brk)}$), 
\beq
{\rm A}^{\rm phys}_{K^+ \to \pi^+ \pi^0} \equiv A_{+0} = {3\over 2}
\left[ A_2^{\rm (true)} + \delta A_2^{\rm (em)} + 
\delta A_2^{\rm (iso-brk)} \right] e^{i\delta_2}
\label{a1}
\eeq
where 
\beq
\delta A_2^{\rm (em)} = - {2 \over 3} \left[ 
{1 \over F_\pi^3} g_{\rm emw} 
+ { g_8 \over F_\pi^3} \delta m_\pi^2  - 
\delta A_2^{\rm (h-o)} \right] \ \ .  
\label{a2}
\eeq

The first term in Eq.~(\ref{a2}) was the subject of the analysis 
in Sect.~III and has been discussed in great detail.  
The next term has a somewhat subtle origin.  
The contribution from the $\Delta I = 1/2$ weak interaction 
${\cal L}_8$ of Eq.~(\ref{12}) to the $K^+ \to \pi^+ \pi^0$ amplitude is 
proportional to $g_8(p_{\pi^+}^2 - p_{\pi^0}^2) = g_8\delta m_\pi^2$. 
This is ordinarily discarded in calculations in 
which isospin is conserved and $\delta m_\pi^2 = 0$. 
It cannot, however, be neglected 
in the present context.  Perhaps the most 
interesting feature of our result is an approximate
cancellation between the first and second terms of Eq.~(\ref{a2}).  
To make this explicit we recall Eq.~(\ref{ld3}) to write 
\beq
g_{\rm emw} = 
 - g_8 \delta m_\pi^2  + \delta g_{\rm emw} \ \ ,
\label{a3}
\eeq
where $\delta g_{\rm emw}$ arises from the sum of the 
intermediate-range part of the chiral contribution and 
the EMP contribution. Our estimate implies 
\beq
{\delta g_{\rm emw} \over g_8 \delta m_\pi^2 } = 0.38 \pm 0.19
\eeq
for this quantity.  

Finally, the contribution $\delta A_2^{\rm (h-o)}$ 
in Eq.~(\ref{a2}) represents electromagnetic 
corrections of higher order in the chiral expansion which vanish 
in the chiral limit. The Feynman diagrams for the photonic corrections
to the $K^+ \to \pi^+\pi^0$ amplitude also generate effects at
order $(e^2p^2)$,  and we find
\beq
\delta A_2^{\rm (h-o)} = {3 \alpha g_8\over 4\pi F_\pi^3} 
\left[  m_\pi^2 \ln \left({\Lambda^2 \over m_\pi^2}\right) 
+ {3\over 2} m_\pi^2 + \dots \right] \ \ .
\label{a4}
\eeq
In $\delta A_2^{\rm (h-o)}$ there is a residual dependence on the 
cutoff $\Lambda$.  However, since it enters only logarithmically 
and is multiplied by a small factor of $m_\pi^2$ it is 
inconsequential for the final answer. We have simply set 
$\Lambda^2 = m_\rho^2$ in this contribution.  In addition, at this
order in the chiral expansion one also needs to
include meson loop effects, involving both the effects of $\delta
m^2_\pi$ and the loops proportional to $g_{emw}$.
This can be done in chiral perturbation theory. Our
evaluation of $g_{emw}$ becomes an input in that calculation and the
photon loop result of Eq.~(\ref{a4}) is relevant for the determination
of the chiral constants at order $(e^2p^2)$. At this stage, we will
include only the effects of Eq.~(\ref{a4}) and reserve a full
calculation at
next order for a future publication~{\cite{cdg}}.

Overall, the net result of our calculation is a shift 
due to electromagnetic effects in the
apparent $A_2$ amplitude ranging over $0.6 \to 2.6\%$ 
depending on how the matching is carried out.   Taking 
the mean value, we obtain 
\beq
{\delta A_2^{\rm (em)} \over A_2} = - 0.016 \pm 0.01 \ \ ,
\label{a6}
\eeq

We do not calculate the contribution $\delta A_2^{\rm (iso-brk)}$ 
in Eq.~(\ref{a1}) which arises from the mixing 
between $\pi^0$ and $\eta, \eta'$, 
\beq
\delta A_2^{\rm (iso-brk)} = \delta A_2^{\pi^0 -\eta} + 
\delta A_2^{\pi^0 -\eta'} \ \ .
\label{a5a}
\eeq
This effect is primarily due to quark mass differences and 
has already been analyzed in Ref.~\cite{dgh2}.  The result 
cited there is\footnote{Note that Eq.~(IX-3.21) of 
Ref.~\cite{dgh2} contains a minus sign typo.}
\beq
{\delta A_2^{\rm (iso-brk)} \over A_2} = 
{\delta A_2^{\pi^0 -\eta} \over A_2} + 
{\delta A_2^{\pi^0 -\eta'} \over A_2} \simeq 0.14 + 0.21 = 0.35 \ .
\label{a5b}
\eeq

\section{Conclusions}
In the $K\to 2 \pi$ amplitudes, the ratio of 
$\Delta I= 1/2$ and $\Delta I= 3/2$ amplitudes is 
about $22$.  This suggests that electromagnetic corrections 
to the former can lead to contributions to the latter of order 
$22/137$ or around $20 \%$.  Indeed, we have found 
individual contributions of order $10 \%$ to occur.  
If added up, these electromagnetic corrections to the 
$K^+ \to \pi^+ \pi^0$ amplitude would contribute at the $20\%$ 
level.  However, there turns out to be 
a significant cancellation which greatly weakens the effect.  
The realization of this cancellation requires a consistent 
application of electromagnetic effects to both the pion masses 
and the weak amplitude. Including all scales in the electromagnetic 
shifts leads to this cancellation only being partial. Due to this
cancellation, we have assigned
a generous fractional uncertainty to the final answer. However in
absolute terms, the overall
electromagnetic effect that we have calculated 
at this order
is only a small part of the experimental value
of $A_2$. Given the
knowledge of the long and short distance components of the amplitude,
our confidence in this result is quite strong.

\vspace{1.6cm}

The research described here was supported in part by the 
National Science Foundation.  One of us (V.C.) acknowledges 
support from M.U.R.S.T.

\eject

\end{document}